\documentstyle[amsfonts,epsfig,portland]{aipproc}

\newcommand{\be}{\begin{equation}}
\newcommand{\ee}{\end{equation}}
\newcommand{\beq}{\begin{eqnarray}}
\newcommand{\eeq}{\end{eqnarray}}

\begin{document}
\title{Topological Defects and Cosmology\thanks{Invited lectures at WHEPP-5,
IUCAA, India, Jan. 12 - 26 1998.}}

\author{Robert H. Brandenberger}
\address{Brown University Physics Department\\
Providence, R.I. 02912, USA}

\maketitle

\begin{abstract}
Many particle physics models of matter admit solutions corresponding to stable or long-lived topological defects. In the context of standard cosmology it is then unavoidable that such defects will form during phase transitions in the very early Universe. Certain types of defects lead to disastrous consequences for cosmology, others may play a useful role, as possible seeds for the formation of structure in the Universe, or in mediating baryon number violating processes. In all cases, topological defects lead to a fruitful interplay between particle physics and cosmology.  
\end{abstract}

\section{Introduction}

Most aspects of high energy physics beyond the standard model can only be tested by going to energies far greater than those which present accelerators can provide. Fortunately, the marriage between particle physics and cosmology has provided a way to ``experimentally" test the new theories of fundamental forces.

The key realization is that physics of the very early Universe may explain the origin of the structure on the scales of galaxies and beyond. It now appears that a rich set of data concerning the nonrandom distribution of matter on a wide range of cosmological scales, and on the anisotropies in the cosmic microwave background (CMB), may potentially be explained by high energy physics. Topological defects provide one class of mechanisms for generating the initial density fluctuations$^{\cite{ZelVil}}$. Topological defects are relevant to cosmology in other ways. Models with defects often predict observational signatures which are not detected. In this way, studying the consequences of particle physics models in the context of cosmology may lead to severe constraints on new microscopic theories. Finally, particle physics and field theory may provide an explanation for another deep cosmological puzzle, namely the origin of the observed small but non-vanishing net baryon to entropy ratio.

\section{Defect Formation and Classification}

According to our current view of particle physics, matter at high energies and
temperatures must be described in terms of fields. Gauge symmetries have
proved to be extremely useful in describing the standard model of particle
physics. Spontaneous breaking of an internal symmetry group is a crucial building block of both the standard model and its extensions such as {\it Grand Unified Theories} and {\it Supersymmetry}. 

Spontaneous symmetry breaking is induced by an order parameter $\varphi$ taking
on a nontrivial expectation value $< \varphi >$ below a certain temperature
$T_c$.  In some particle physics models, $\varphi$ is a fundamental scalar
field in a nontrivial representation of the gauge group $G$ which is broken.
However, $\varphi$ could also be a fermion condensate, as in the BCS theory of superconductivity. In many models, topological defects are unavoidable products of the symmetry breaking phase transition.
  
Consider a single component real scalar field with a typical symmetry breaking
potential
\be \label{stringpot}
V (\varphi) = {1\over 4} \lambda (\varphi^2 - \eta^2)^2 \, ,  
\ee
where $\lambda \ll 1$ is a coupling constant. In the presence of a thermal heat bath at a temperature $T$, finite temperature effects lead to a correction to the free energy density (\ref{stringpot}) which are (to leading order) proportional to $T^2 \varphi^2$. At temperatures above a critical temperature $T_c \sim \eta$, the symmetry is unbroken and the energetically favored state is $\varphi = 0$, whereas for temperatures below $T_c$, the state $\varphi = 0$ becomes unstable and the symmetry is broken. The state which now minimizes the free energy density is not unique. The states minimizing $V(\varphi)$ form the {\it vacuum manifold} ${\cal M}$, in our example $\varphi = \pm \eta$.

The phase transition will take place over a time interval short compared to the expansion time of the Universe,  and will lead to correlation regions of radius $\xi < t$ inside of which $\varphi$ is approximately constant, but outside of which $\varphi$ ranges randomly over ${\cal M}$. The correlation regions are separated by topological defects, in the above example domain walls, regions in
space where $\varphi$ leaves the vacuum manifold ${\cal M}$ and where,
therefore, potential energy is localized.  

Topological defects are familiar from solid state and condensed matter
systems.  Crystal defects, for example, form when water freezes or
when a metal crystallizes.  Point defects, line defects and planar
defects are possible.  Defects are also common in liquid crystals$^{\cite{LQ}}$.
They arise in a temperature quench from the disordered to the ordered
phase.  Vortices in $^4$He are analogs of global cosmic strings.
Vortices and other defects are also produced$^{\cite{Salomaa}}$ during a quench below the critical temperature in $^3$He.  Finally, vortex lines may play an
important role in the theory of superconductivity$^{\cite{Abrikosov}}$.

The analogies between defects in particle physics and condensed matter
physics are quite deep.  Defects form for the same reason: the vacuum
manifold is topologically nontrivial.  The arguments$^{\cite{Kibble1}}$ which say that in
a theory which admits defects, such defects will inevitably form, are
applicable both in cosmology and in condensed matter physics.

The symmetry breaking phase transition takes place at $T = T_c$. From condensed matter physics it is well
known that in many cases topological defects form during phase transitions,
particularly if the transition rate is fast on a scale compared to the system
size.  When cooling a metal, defects in the crystal configuration will be
frozen in; during a temperature quench of $^4$He, thin vortex tubes of the
normal phase are trapped in the superfluid; and analogously in a temperature
quench of a superconductor, flux lines are trapped in a surrounding sea of the
superconducting Meissner phase.

In cosmology, the rate at which the phase transition proceeds is given by the
expansion rate of the Universe which is very fast in the early Universe.  Hence, topological defects will inevitably be
produced in a cosmological phase transition$^{\cite{Kibble1}}$, provided the underlying particle physics model allows such defects. 
 
The argument which  ensures that in theories which admit
topological defects, such defects will be produced
during a phase transition in the very early Universe is called the Kibble mechanism$^{\cite{Kibble1}}$. At high temperatures $T \gg T_c$, $\varphi = 0$.  When $T$ drops below $T_c$, the point $\varphi = 0$ becomes unstable and at all points in space the field will start rolling towards the vacuum manifold. However, the fluctuations which determine in which direction the field rolls at points separated by a large distance $s$ are uncorrelated.
For a system in thermal equilibrium, the correlation length $\xi (t)$, the length $s$ beyond which
the fluctuations are random, is bounded from above by causality:
\be
\xi (t) < t \, .\ee
 
We now turn to the classification of
topological defects$^{\cite{Kibble1}}$.  We consider theories with an $n$-component order
parameter $\varphi$ and with a free energy
density of the form (\ref{stringpot}) with $\varphi^2 = \sum\limits^n_{i = 1} \, \varphi^2_i$. 
There are various types of local and global topological defects
(regions of trapped energy density) depending on the number $n$ of components
of $\varphi$ (see e.g. \cite{VilShell} for a comprehensive survey of topological defect models). The words ``local" and ``global" refer to whether the symmetry
which is broken is a gauge or global symmetry.  In the case of local
symmetries, the topological defects have a well defined core outside of which
$\varphi$ contains no energy density in spite of non-vanishing gradients
$\nabla \varphi$:  the gauge fields $A_\mu$ can absorb the gradient,
{\it i.e.,} $D_\mu \varphi = 0$ when $\partial_\mu \varphi \neq 0$,
where the covariant derivative $D_\mu$ is defined by
$D_\mu = \partial_\mu + ie \, A_\mu$, $e$ being the gauge coupling constant.
Global topological defects, however, have long range density fields and
forces.
\par
For $n = 1$, the vacuum manifold (set of ground states) consists of two points.
The defects which result are two dimensional, domain walls. For $n = 2$, the
vacuum manifold is a circle. This leads to one dimensional defects, cosmic
strings. If $n = 3$, then the set of ground states is $S^2$ and the resulting
defects are monopoles. Obviously, in three space dimensions the dimensionality
of the defects is $3 - n$. For $n = 4$ no stable defects exist. There are,
however, space-time defects called textures.


   
Theories with domain walls are ruled out$^{\cite{nodw}}$ since a single domain wall stretching
across the Universe today would over-close the Universe.  Local monopoles are
also ruled out$^{\cite{nomon}}$ since they would over-close the Universe.  Local
textures are ineffective at producing structures because there is no trapped potential energy, and since the spatial gradients of the condensate can be compensated by gauge fields.

As an example, let us consider local cosmic strings (see e.g. \cite{VilShell,HK95,RB94} for recent reviews). These arise
in theories with a complex order parameter $(n = 2)$. In this case the vacuum manifold of the model is ${\cal M} = S^1$ which has non-vanishing first homotopy group:
\be
\Pi_1 ({\cal M}) = Z \neq 1 \, . 
\ee
A cosmic string is a line of trapped energy density which arises
whenever the field $\varphi (x)$ circles ${\cal M}$ along a closed path
in space ({\it e.g.}, along a circle).  In this case, continuity of
$\varphi$ implies that there must be a point with $\varphi = 0$ on any
disk whose boundary is the closed path.  

To construct a field configuration with a string along the $z$ axis$^{\cite{Nielsen}}$,
take $\varphi (x)$ to cover ${\cal M}$ along a circle with radius $r$
about the point $(x,y) = (0,0)$:
\be \label{stringconf}
\varphi (r, \vartheta ) \simeq \eta e^{i \vartheta} \, , \, r \gg
\eta^{-1} \, .  
\ee
This configuration has winding number 1, {\it i.e.}, it covers ${\cal
M}$ exactly once.  Maintaining cylindrical symmetry, we can extend
(\ref{stringconf}) to arbitrary $r$
\be
\varphi (r, \, \vartheta) = f (r) e^{i \vartheta} \, ,  
\ee
where $f (0) = 0$ and $f (r)$ tends to $\eta$ for large $r$.  The
width $w$ can be found by balancing potential and tension
energy.  The result is 
\be
w \, \sim \, \lambda^{-1/2} \eta^{-1} \, .
\ee

For local cosmic strings, the energy density decays
exponentially for $r \gg w$.  In this case, the energy $\mu$
per unit length of a string is finite and depends only on the symmetry
breaking scale $\eta$ 
\be
\mu \sim \eta^2  
\ee
(independent of the coupling $\lambda$).  The value of $\mu$ is the
only free parameter in a cosmic string model.

Monopoles are point-like defects which can be constructed in a way similar to the above, and which arise in models with $\Pi_2({\cal M}) \neq 1$, e.g. in our toy model with $n = 3$. Textures$^{\cite{Turok89}}$ arise in models with non-vanishing $\Pi_3({\cal M})$, e.g. in a toy model like (\ref{stringpot}) with $n = 4$. In contrast to the case of domain walls, strings and monopoles, textures contain no trapped potential energy: a texture is a configuration of spatial gradient energy which is unstable. Texture configurations with Hubble radius spatial extent will collapse within one Hubble expansion time. This is, however, long enough for them to seed inhomogeneities in the early Universe.
 
We conclude this section with a brief discussion of the evolution of a cosmic string network for $t > t_G$. The key processes are loop production
by intersections of infinite strings and loop shrinking
by gravitational radiation.  These two processes combine to create a
mechanism by which the infinite string network loses energy (and
length as measured in comoving coordinates).  It can be shown (see e.g. \cite{Vil85}) that, as a consequence, the correlation length of the string network is always proportional to its causality limit $\xi (t) \sim t$. Hence, the energy density $\rho_\infty (t)$ in long strings is a fixed
fraction of the background energy density $\rho_c (t)$
\be
\rho_\infty (t) \sim \mu \xi (t)^{-2} \sim \mu t^{-2}  
\ee
or
\be
{\rho_\infty (t)\over{\rho_c (t)}} \sim G \mu \, . 
\ee

We conclude that the cosmic string network approaches a ``scaling
solution" in which the statistical properties of the
network are time independent if all distances are scaled to the
horizon distance.

\section{Defects and Structure Formation}

The only local field theory defects which could be relevant for structure formation are cosmic strings. Hence, this section will focus on the cosmic string theory of structure formation. The starting point in this theory is the scaling solution for the cosmic string network,
according to which at all times $t$ (in particular at $t_{eq}$, the
time when perturbations can start to grow) there will be a few long
strings crossing each Hubble volume, plus a distribution of loops of
radius $R \ll t$. 

The cosmic string model admits three mechanisms for structure
formation:  loops, filaments, and wakes.  Cosmic string loops have the same
time averaged field as a point source with mass$^{\cite{Turok84}}$
$ M (R) = \beta R \mu $, $R$ being the loop radius and $\beta \sim 2 \pi$.  Hence, loops will be seeds for spherical accretion of dust and radiation.
However, according to cosmic string evolution simulations$^{\cite{CSsimuls}}$, most of the energy in strings is in the long strings, and hence the loop accretion mechanism is sub-dominant. 

The second mechanism involves long strings moving with relativistic
speed in their normal plane which give rise to
velocity perturbations in their wake$^{\cite{SilkVil}}$. Space normal to the string is a cone with deficit angle$^{\cite{Vil81}}$
\be \label{deficit}
\alpha = 8 \pi G \mu \, . 
\ee
If the string is moving with normal velocity $v$ through a bath of dark
matter, a velocity perturbation
\be
\delta v = 4 \pi G \mu v \gamma  
\ee
[with $\gamma = (1 - v^2)^{-1/2}$] towards the plane behind the string
results.  At times after $t_{eq}$, this induces planar overdensities (``wakes"),
the most prominent ({\it i.e.}, thickest at the present time) and numerous of which were created at $t_{eq}$, the time of equal matter and
radiation$^{\cite{TV86,AS87,LP90}}$.  The
corresponding planar dimensions are proportional to the Hubble radius at $t_{eq}$ in comoving coordinates.
   
The thickness $d$ of these wakes can be calculated using the
Zel'dovich approximation$^{\cite{Zeld70}}$.  The result is (for $G \mu = 10^{-6}$)
\be
d \simeq G \mu v \gamma (v) z (t_{eq})^2 \, t_{eq} \simeq 4 v \, {\rm
Mpc} \, . 
\ee
 
Wakes arise if there is little small scale structure on the string.
In this case, the string tension equals the mass density, the string
moves at relativistic speeds, and there is no local gravitational
attraction towards the string.

In contrast, if there is small scale structure on strings,
then the coarse-grained string tension $T$ is smaller$^{\cite{Carter}}$ 
than the mass per unit length $\mu$ , and thus there
is a gravitational force towards the string which gives rise to
cylindrical accretion, producing filaments$^{\cite{fils}}$.

Which of the mechanisms -- filaments or wakes -- dominates is
determined by the competition between the velocity induced by the Newtonian gravitational potential of the strings and the velocity perturbation of the wake.   

The cosmic string model predicts a scale-invariant spectrum of density
perturbations, exactly like inflationary Universe models but for a
rather different reason.  Consider the {\it r.m.s.} mass fluctuations
on a length scale $2 \pi k^{-1}$ at the time $t_H (k)$ when this scale
enters the Hubble radius.  From the cosmic string scaling solution it
follows that a fixed ({\it i.e.}, $t_H (k)$ independent) number
$\tilde v$ of strings of length of the order $t_H (k)$ contribute to
the mass excess $\delta M (k, \, t_H (k))$.  Thus
\be
{\delta M\over M} \, (k, \, t_H (k)) \sim \, {\tilde v \mu t_H
(k)\over{G^{-1} t^{-2}_H (k) t^3_H (k)}} \sim \tilde v \, G \mu \, .
\ee
Note that the above argument predicting a scale invariant spectrum
will hold for all topological defect models which have a scaling
solution, in particular also for global monopoles and textures.

The amplitude of the {\it r.m.s.} mass fluctuations (equivalently: of
the power spectrum) can be used$^{\cite{TB86}}$ to normalize $G \mu$.  Since today on
galaxy cluster scales
\be
{\delta M\over M} (k, \, t_0) \sim 1 \, ,  
\ee
the growth rate of fluctuations linear in $a(t)$ yields
\be
{\delta M\over M} \, (k, \, t_{eq}) \sim 10^{-4} \, ,  
\ee
and therefore, using $\tilde v \sim 10$,
\be
G \mu \sim 10^{-5} \, .  
\ee
A similar value is obtained by normalizing the model to the COBE amplitude of CMB anisotropies on large angular scales$^{\cite{BBS,Periv}}$ (the normalizations from COBE and from the power spectrum of density perturbations on large scales agree to within a factor of 2).
Thus, if cosmic strings are to be relevant for structure formation,
they must arise due to a symmetry breaking at energy scale $\eta
\simeq 10^{16}$GeV.  This scale happens to be the scale of unification (GUT)
of weak, strong and electromagnetic interactions.  It is tantalizing
to speculate that cosmology is telling us that there indeed was new
physics at the GUT scale.

With cosmic strings, hot dark matter (HDM) is viable, in contrast to the situation in inflationary Universe models where the adiabatic fluctuations on scales of galaxies get erased by free streaming.  Cosmic string
loops survive free streaming, and can
generate nonlinear structures on galactic scales, as discussed in
detail in \cite{BKST,Edbert}.  Accretion of hot dark matter by a string wake
was studied in \cite{LP90}. In this case, nonlinear perturbations
develop only late. Accretion onto loops and small scale structure on the long strings provide two mechanisms which may lead to high redshift objects such as quasars and high redshift galaxies. The first mechanism has recently been studied in \cite{MB96}, the second in \cite{AB95,ZLB96}.

The power spectrum of density fluctuations in a cosmic string model with HDM has recently been studied numerically by M\"ah\"onen$^{\cite{Mahonen}}$, based on previous work of \cite{Hara} (see also \cite{AS92} for an earlier semi-analytical study). The spectral shape agrees quite well with observations.   
Over the past year there has been significant progress in our ability to calculate the angular power spectrum of CMB anisotropies and the linear power spectrum of density perturbations in defect models$^{\cite{PST}}$ using the same simulations. This now allows us to quantify the bias problem which arises in defect models: since the primordial perturbations are isocurvature and not adiabatic, a COBE-normalized defect model generically predicts insufficient linear power in the spectrum of density perturbations to match the observations$^{\cite{PST,ADSS,ABR}}$. The required bias factor depends, however, quite sensitively on the details of the defect scaling solution which are at present not well understood. With strings and hot dark matter, a bias factor of between 2 or 3$^{\cite{AAS}}$ seems to be required (although the work of \cite{Mahonen} gives a smaller bias). It would be premature to conclude, however, that defect models are inconsistent with observations. The {\it standard} inflationary cold dark matter model also has a bias problem (in this case a COBE-normalized model predicts too much linear power), a problem which
can be at least partially solved by going to an open Universe or adding a cosmological constant. In a similar way, the bias problem for cosmic strings can be substantially lessened$^{\cite{AAS,ABR2}}$. It is, however, also important to keep in mind that a cosmic string model with hot dark matter in a natural way gives rise to a bias factor greater than 1.    

A more robust way of differentiating between defect models and adiabatic inflationary theories is to focus on predictions which are truly distinctive.
The cosmic string theory of structure formation makes several such predictions, both in terms of the galaxy distribution and in terms of CMB anisotropies. On large scales (corresponding to the comoving Hubble radius at $t_{eq}$ and larger, structure is predicted to be dominated either by planar$^{\cite{TV86,AS87,LP90}}$ or filamentary$^{\cite{fils}}$ galaxy concentrations. For models in which the strings have no local gravity, the resulting nonlinear structures will look very different from the nonlinear structures in models in which local gravity is the dominant force. As discovered and discussed recently in \cite{SB96}, a baryon number excess is predicted in the nonlinear wakes. This may explain the ``cluster baryon crisis"$^{\cite{clusterbaryon}}$, the fact that the ratio of baryons to dark matter in rich clusters is larger than what is compatible with the nucleosynthesis constraints in a spatially flat Universe. 

As described earlier, space perpendicular to a long straight
cosmic string is conical with deficit angle given by (\ref{deficit}).  Consider
now CMB radiation approaching an observer in a direction normal to the
plane spanned by the string and its velocity vector.
Photons arriving at the observer having passed on different sides of
the string will obtain a relative Doppler shift which translates into
a temperature discontinuity of amplitude$^{\cite{KS84}}$
\be
{\delta T\over T} = 4 \pi G \mu v \gamma (v) \, ,  
\ee
where $v$ is the velocity of the string.  Thus, the distinctive
signature for cosmic strings in the microwave sky are line
discontinuities in $T$ of the above magnitude.

Given ideal maps of the CMB sky it would be easy to detect strings.
However, real experiments have finite beam width.  Taking into account
averaging over a scale corresponding to the beam width will smear out
the discontinuities, and it turns out to be surprisingly hard to
distinguish the predictions of the cosmic string model from that of
inflation-based theories using quantitative statistics which are easy
to evaluate analytically, such as the kurtosis of the spatial gradient
map of the CMB$^{\cite{MPB94}}$ (see also \cite{Leandros,AM} for other recently suggested statistics and for further references). The angular power spectrum of the CMB on scales of a degree or less can also be used to discriminate between different models. Whereas inflation-based adiabatic theories predict a sequence of fairly narrow {\it Doppler} peaks, defect models do not predict secondary peaks, and the first peak is often quite broad$^{\cite{AAcrew,HuWhite,AAS,ABR2}}$, the basic reason being because the tensor and vector modes are important (see also \cite{RAT} for first evidence of the importance of tensor models). Forthcoming experiments will easily be able to discriminate between the various predictions.

Global textures also produce distinctive non-Gaussian signatures$^{\cite{TurSper}}$ in CMB maps. In fact, these signatures are more pronounced and on larger scales than the signatures in the cosmic string model.

\section{Defects and Baryogenesis}

Baryogenesis is another area where particle physics and cosmology connect in a very deep way. The goal is to explain the observed asymmetry between matter and antimatter in the Universe. In particular, the objective is to be able to explain the observed value of the net baryon to entropy ratio at the present time
\be
{{\Delta n_B} \over s}(t_0) \, \sim \, 10^{-10} 
\ee
starting from initial conditions in the very early Universe when this ratio vanishes. Here, $\Delta n_B$ is the net baryon number density and $s$ the entropy density.

As pointed out by Sakharov$^{\cite{Sakharov}}$, three basic criteria must be satisfied in order to have a chance at explaining the data:
The theory describing the microphysics must contain baryon number violating processes, these processes must be C and CP violating, and
they must occur out of thermal equilibrium.

As was discovered in the 1970's$^{\cite{GUTBG}}$, all three criteria can be satisfied in GUT theories. In these models, baryon number violating processes are mediated by superheavy Higgs and gauge particles. The baryon number violation is visible in the Lagrangian, and occurs in perturbation theory (and is therefore in principle easy to calculate). In addition to standard model CP violation, there are typically many new sources of CP violation in the GUT sector. The third Sakharov condition can also be realized: After the GUT symmetry-breaking phase transition, the superheavy particles may fall out of thermal equilibrium. The out-of-equilibrium decay of these particles can thus generate a non-vanishing baryon to entropy ratio. 

The magnitude of the predicted $n_B / s$ depends on the asymmetry $\varepsilon$ per decay, on the coupling constant $\lambda$ of the $n_B$ violating processes, and on the ratio $n_X / s$ of the number density $n_X$ of superheavy Higgs and gauge particles to the number density of photons, evaluated at the time $t_d$ when the baryon number violating processes fall out of thermal equilibrium, and assuming
that this time occurs after the phase transition. The quantity $\varepsilon$ is proportional to the CP-violation parameter in the model. In a GUT theory, this CP violation parameter can be large (order 1), whereas in the standard electroweak theory it is given by the CP violating phases in the CKM mass matrix and is very small. As shown in \cite{GUTBG} it is easily possible to construct models which give the right $n_B / s$ ratio after the GUT phase transition (for recent reviews of baryogenesis see \cite{Dolgov} and \cite{RubShap}).
 
The ratio $n_B / s$, however, does not only depend on $\varepsilon$, but also on $n_X / s (t_d)$. If the temperature $T_d$ at the time $t_d$ is greater than the mass $m_X$ of the superheavy particles, then it follows from the thermal history in standard cosmology that $n_X \sim s$. However, if $T_d < m_X$, then the number density of $X$ particles is diluted exponentially in the time interval between when $T = m_X$ and when $T = T_d$. Thus, the predicted baryon to entropy ratio is exponentially suppressed:
\be \label{expdecay}
{n_B \over s} \, \sim \, {1 \over {g^*}} \lambda^2 \varepsilon e^{- m_X / T_d} \, ,
\ee
where $g^*$ is the number of spin degrees of freedom in thermal equilibrium at the time of the phase transition.
In this case, the standard GUT baryogenesis mechanism is ineffective.

However, topological defects may come to the rescue$^{\cite{BDH92}}$. As we have seen in the previous section, topological defects will inevitably be produced in the symmetry breaking GUT transition provided they are topologically allowed in that symmetry breaking scheme. The topological defects provide an alternative mechanism of GUT baryogenesis.

Inside of topological defects, the GUT symmetry is restored. In fact, the defects can be viewed as solitonic configurations of $X$ particles. The continuous decay of defects at times after $t_d$ provides an alternative way to generate a non-vanishing baryon to entropy ratio. The defects constitute out of equilibrium configurations, and hence their decay can produce a non-vanishing $n_B / s$ in the same way as the decay of free $X$ quanta. 

The way to compute the estimate $n_B / s$ ratio is as follows: The defect
scaling solution gives the energy density in defects at all times. Taking the time derivative of this density, and taking into account the expansion of the Universe, we obtain the loss of energy attributed to defect decay. By energetics, we can estimate the number of decays of individual quanta which the defect decay corresponds to. We can then use the usual perturbative results to compute the resulting net baryon number.

Provided that $m_X < T_d$, then at the time when the baryon number violating processes fall out of equilibrium (when we start generating a non-vanishing $n_B$) the energy density in free $X$ quanta is much larger than the defect density, and hence the defect-driven baryogenesis mechanism is subdominant. However, if $m_X > T_d$, then as indicated in (\ref{expdecay}), the energy density in free quanta decays exponentially. In contrast, the density in defects only
decreases as a power of time, and hence soon dominates baryogenesis.

One of the most important ingredients in the calculation is the time dependence of $\xi(t)$, the separation between defects. Immediately after the phase transition at the time $t_f$ of the formation of the defect network, the separation is $\xi(t_f) \sim \lambda^{-1} \eta^{-1}$. In the time period immediately following, the time period of relevance for baryogenesis, $\xi(t)$ approaches the Hubble radius according to the equation$^{\cite{Kibble2}}$
\be \label{defsep}
\xi(t) \, \simeq \, \xi(t_f) ({t \over {t_f}})^{5/4} \, .
\ee
Using this result to calculate the defect density, we obtain after some algebra
\be \label{barres}
{{n_B} \over s}|_{\rm defect} \, \sim \, \lambda^2 {{T_d} \over \eta} {{n_B} \over s}|_0 \, ,
\ee
where $n_B / s|_0$ is the unsuppressed value of $n_B / s$ which can be obtained using the standard GUT baryogenesis mechanism. We see from (\ref{barres}) that even for low values of $T_d$, the magnitude of $n_B / s$ which is obtained via the defect mechanism is only suppressed by a power of $T_d$. However, the maximum strength of the defect channel is smaller than the maximum strength of the usual mechanism by a geometrical suppression factor $\lambda^2$ which expresses the fact that even at the time of defect formation, the defect network only occupies a small volume.

It has been known for some time that there are baryon number violating processes even in the standard electroweak theory. These processes are, however, nonperturbative. They are connected with the t'Hooft anomaly$^{\cite{tHooft}}$, which in turn is due to the fact that the gauge theory vacuum is degenerate, and that the different degenerate vacuum states have different quantum numbers (Chern-Simons numbers). In theories with fermions, this implies different baryon number. Configurations such as sphalerons$^{\cite{sphal}}$ which interpolate between two such vacuum states thus correspond to baryon number violating processes.

As pointed out in \cite{KRS85}, the anomalous baryon number violating processes are in thermal equilibrium above the electroweak symmetry breaking scale. Therefore, any net baryon to entropy ratio generated at a higher scale will be erased, unless this ratio is protected by an additional quantum number such as a non-vanishing $B - L$ which is conserved by electroweak processes.

However, as first suggested in \cite{Shap} and discussed in detail in many recent papers (see \cite{EWBGrev} for reviews of the literature), it is possible to regenerate a non-vanishing $n_B / s$ below the electroweak symmetry breaking scale. Since there are $n_B$ violating processes and both C and CP violation in the standard model, Sakharov's conditions are satisfied provided that one can realize an out-of-equilibrium state after the phase transition. Standard model CP violation is extremely weak. Thus, it appears necessary to add some sector with extra CP violation to the standard model in order to obtain an appreciable $n_B / s$ ratio. A simple possibility which has been invoked often is to add a second Higgs doublet to the theory, with CP violating relative phases. 

The standard way to obtain out-of-equilibrium baryon number violating processes immediately after the electroweak phase transition is$^{\cite{EWBGrev}}$ to assume that the transition is strongly first order and proceeds by the nucleation of bubbles (note that these are two assumptions, the second being stronger than the first!). 

Bubbles are out-of-equilibrium configurations. Outside of the bubble (in the false vacuum), the baryon number violating processes are unsuppressed, inside they are exponentially suppressed. In the bubble wall, the Higgs fields have a nontrivial profile, and hence (in models with additional CP violation in the Higgs sector) there is enhanced CP violation in the bubble wall. In order to obtain net baryon production, one may either use fermion scattering off bubble walls$^{\cite{CKN1}}$ (because of the CP violation in the scattering, this generates a lepton asymmetry outside the bubble which converts via sphalerons to a baryon asymmetry) or sphaleron processes in the bubble wall itself$^{\cite{TZ,CKN2}}$. It has been shown that, using optimistic parameters (in particular a large CP violating phase $\Delta \theta_{CP}$ in the Higgs sector) it is possible to generate the observed $n_B / s$ ratio. The resulting baryon to entropy ratio is of the order
\be \label{ewres}
{{n_B} \over s} \, \sim \, \alpha_W^2 (g^*)^{-1} \bigl( {{m_t} \over T} \bigr)^2 \Delta \theta_{CP} \, ,
\ee
where $\alpha_W$ refers to the electroweak interaction strength, $g^*$ is the number of spin degrees of freedom in thermal equilibrium at the time of the phase transition, and $m_t$ is the top quark mass. The dependence on the top quark mass enters because net baryogenesis only appears at the one-loop level.

However, analytical and numerical studies show that, for the large Higgs masses which are indicated by the current experimental bounds, the electroweak phase transition will unlikely be sufficiently strongly first order to proceed by bubble nucleation. In addition, there are some concerns as to whether it will proceed by bubble nucleation at all (see e.g. \cite{Gleiser,Gavai}).

Once again, topological defects come to the rescue. In models which admit defects, such defects will inevitably be produced in a phase transition independent of its order. Moving topological defects can play the same
role in baryogenesis as nucleating bubbles. In the defect core, the electroweak symmetry is unbroken and hence sphaleron processes are unsuppressed$^{\cite{Perkins}}$. In the defect walls there is enhanced CP violation for the same reason as in bubble walls. Hence, at a fixed point in space, a non-vanishing baryon number will be produced when a topological defect passes by.

Defect-mediated electroweak baryogenesis has been worked out in detail in \cite{BDPT} (see \cite{BDT} for previous work) in the case of cosmic strings. The scenario is as follows: at a particular point $x$ in space, antibaryons are produced when the front side of the defect passes by. While $x$ is in the defect core, partial equilibration of $n_B$ takes place via sphaleron processes. As the back side of the defect passes by, the same number of baryons are produced as the number of antibaryons when the front side of the defect passes by. Thus, at the end a positive number of baryons are left behind.

As in the case of defect-mediated GUT baryogenesis, the strength of defect-mediated electroweak baryogenesis is suppressed by the ratio ${\rm SF}$ of the volume which is passed by defects divided by the total volume, i.e.
\be
{{n_B} \over s} \, \sim \, {\rm SF} {{n_B} \over s}|_0 \, ,
\ee
where $(n_B / s)|_0$ is the result of (\ref{ewres}) obtained in the bubble nucleation mechanism. 

A big caveat for defect-mediated electroweak baryogenesis is that the standard electroweak theory does not admit topological defects. However, in a theory with additional physics just above the electroweak scale it is possible to obtain defects (see e.g. \cite{TDB95,RB98} for some specific models). The closer the scale $\eta$ of the new physics is to the electroweak scale $\eta_{EW}$, the larger the volume in defects and the more efficient defect-mediated electroweak baryogenesis. Using the result of (\ref{defsep}) for the separation of defects, we obtain (for non-superconducting strings)
\be
{\rm SF} \, \sim \, \lambda \bigl( {{\eta_{EW}} \over \eta} \bigr)^{3/2} v_D\, .
\ee 
where $v_D$ is the mean defect velocity.
 
Obviously, the advantage of the defect-mediated baryogenesis scenario is that it does not depend on the order and on the detailed dynamics of the electroweak phase transition.  

\section{Summary}

As we have seen, topological defects may play an important role in cosmology.
Defects are inevitably produced during symmetry breaking phase transitions in the early Universe in all theories in which defects are topologically stable.
Theories giving rise to domain walls or local monopoles are ruled out by cosmological constraints. Those producing cosmic strings, global monopoles and textures are quite attractive.

If the scale of symmetry breaking at which the defects are produced is about $10^{16}$ GeV, then defects can act as the seeds for galaxy formation. Defect theories of structure formation predict a roughly scale-invariant spectrum of density perturbations, similar to inflation-based models. However, the phases in the density field are distributed in a non-Gaussian manner, thus leading to distinctive signatures both in CMB maps and in large-scale structure surveys by means of which the predictions of defect models can be distinguished between each other as well as from those of inflationary models. 

Since defect models lead to isocurvature perturbations on scales larger than the Hubble radius, the relative normalization of CMB anisotropies and of the power spectrum of density perturbations is different from what is obtained in inflation-based adiabatic models. It appears that a bias factor greater than 1 is required in order for defect models to match the current observations. Note, however, that bias is automatic in a cosmic string model with hot dark matter.
The predictions of defect models also differ from those of inflation-based models with respect to the acoustic oscillations in the CMB angular power spectrum. Defect models do not predict the narrow Doppler peaks which adiabatic models predict. Thus, the predictions of defect models can be tested using future CMB experiments.

Topological defects may also play a crucial role in baryogenesis. This applies
both to GUT and electroweak baryogenesis. The crucial point is that defects constitute out-of-equilibrium configurations, and may therefore be the sites of net baryon production.

\medskip
\centerline{\bf Acknowledgments}
\medskip

I wish to thank the organizers of this stimulating workshop for inviting me to
speak in Pune and for their wonderful hospitality. I am grateful to all of my research collaborators, on whose work I have freely drawn. Partial financial support for the preparation of this manuscript has been provided at Brown by the US Department of Energy under Grant DE-FG0291ER40688, Task A.

\end{document}